\documentclass[aps,pra,nobalancelastpage,twocolumn,superscriptaddress,nofootinbib,amsmath,groupedaddress]{revtex4-1}

\usepackage[english]{babel}
\usepackage{CJKutf8}
\usepackage{graphicx}
\usepackage{subfigure}
\usepackage{dcolumn}
\usepackage{bm}
\usepackage{siunitx}
\usepackage{color}
\usepackage{longtable}
\usepackage{amsmath}
\usepackage{amssymb}
\usepackage{array}
\usepackage{multirow}
\usepackage{tablefootnote}
\usepackage{upgreek}
\usepackage[letterpaper,total={7in,9.5in},top=0.75in,left=0.75in]{geometry}
\usepackage[colorlinks=true,linkcolor=blue,citecolor=blue,urlcolor=blue,filecolor=blue]{hyperref}

\definecolor{mypink1}{rgb}{0.858, 0.188, 0.478}
\DeclareSIUnit\gauss{G}

\newcolumntype{L}[1]{>{\raggedright\let\newline\\\arraybackslash\hspace{0pt}}m{#1}}
\newcolumntype{C}[1]{>{\centering\let\newline\\\arraybackslash\hspace{0pt}}m{#1}}
\newcolumntype{R}[1]{>{\raggedleft\let\newline\\\arraybackslash\hspace{0pt}}m{#1}}

\begin{document}

\title{Continuous guided strontium beam with high phase-space density}
\author{Chun-Chia Chen (陳俊嘉)}
\email[]{beam@strontiumBEC.com}
\author{Shayne Bennetts}
\author{Rodrigo Gonz{\'a}lez Escudero}
\author{Benjamin Pasquiou}
\author{Florian Schreck}

\affiliation{Van der Waals-Zeeman Institute, Institute of Physics, University of Amsterdam, Science Park 904,
1098XH Amsterdam, The Netherlands}

\date{\today}

\begin{abstract}
A continuous guided atomic beam of $^{88}\mathrm{Sr}$ with a phase-space density exceeding $10^{-4}$ in the moving frame and a flux of $\num{3e7}  \, \mathrm{at}~\si{\per\second}$ is demonstrated. This phase-space density is around three orders of magnitude higher than previously reported for steady-state atomic beams. We detail the architecture necessary to produce this ultracold atom source and characterize its output after $\sim \SI{4}{\centi \meter}$ of propagation. With radial temperatures of less than $\SI{1}{\micro\kelvin}$ and a velocity of $\SI{8.4}{\cm\per\second}$ this source is ideal for a range of applications. For example, it could be used to replenish the gain medium of an active optical superradiant clock or be employed to overcome the Dick effect that can limit the performance of pulsed-mode atom interferometers, atomic clocks and ultracold atom based sensors in general. Finally, this result represents a significant step towards the development of a steady-state atom laser.
\end{abstract}

\begin{CJK*}{UTF8}{min}

\maketitle

\end{CJK*}

\section{Introduction}
\label{Sec:Introduction}

From atomic clocks \cite{LudLow2015ReviewAtomicClock} to atom interferometers \cite{Cronin2009ReviewAtomInterferometry} cold and ultracold atom devices are defining the state of the art of precision measurement. Cold atom sensors are tackling fundamental questions like detecting dark matter or dark energy \cite{Wcislo2018BoundDarkMatterCoupling, Hees2016DarkMattFountain, Geraci2016DarkMatterAI, Jaffe2017SubGravForces}, gravitational waves \cite{Canuel2018MIGA, Rajendran2013GWatomicsensors, Hogan2011GWsensorAGISLEO, Kolkowitz2016GraveWaveClock} and variations of fundamental constants \cite{Martins2017RevFundamentalConstant, Godun2014ConstraintFundamentalConstant}, as well as making precision measurements of physical constants \cite{Parker2018FineStructureconstant, Bertoldi2006NewtonG, Fixler2007NewtonGconstant, Rosi2014NewtonG}. In the applied domain, optical atomic clocks continue to set new records in timekeeping \cite{Campbell2017FermiDegClock, McGrew2018ClockGeodesy, Nemitz2016RatioYbSrClock}, while cold atom gravimeters, gravity gradiometers, gyroscopes and accelerometers are of growing importance for geology and navigation \cite{Bidel2018MarineGravimetry, Menoret2018TransportableGravimeter, Dutta2016ContinuousInertialSensor, Cheiney2018ContinuousAccelero}. Yet almost all these cold atom sensors and the atom sources they rely on operate in pulsed mode, which poses a fundamental limitation. The Dick effect \cite{Dick1987OriginalDickEffect, Quessada2003DickEffect}, where frequency noise aliasing arises from the dead time between sample interrogations, intrinsically limits the performance of a pulsed device.

Atomic clocks now reach sensitivities where the Dick effect limits performance \cite{Takamoto2011BeyondDick, Al-Masoudi2015NoiseInClock}. Improvements in the optical clock local oscillator has allowed them to better preserve phase across the dead time \cite{Kessler2012SiliconOptCavity, Haefner2015Cavity50cm}. Others synchronize multiple copies of the same apparatus to avoid dead time \cite{Biedermann2013FirstInterleaveClock, Schioppo2016DualYbClock, Dutta2016ContinuousInertialSensor}, or increase the duty cycle by performing multiple measurements after a single sample preparation phase \cite{Kohlhaas2015PhaseLockClockToAtom, Norcia2019TweezerClock, Westergaard2010MinimizingDickEffect}. Hybridization of a cold atom interferometer with other devices can combine the low offsets of atom interferometers with the higher bandwidth of classical devices in a single apparatus \cite{Cheiney2018ContinuousAccelero}. A fundamentally simpler approach would be to create a fully continuous device \cite{Keith1991AtomInterf, Durfee2006ReversibleSagnacGyro, Jallageas2018FoCScontinuousFountain, Xue2015ContinuousColdAI}. Active optical clocks \cite{Chen2005ActiveClock, Meiser2009MilliHzLaser} are a promising proposal for producing a new generation of optical clocks that is inherently continuous, circumventing both the Dick effect as well as other challenges now limiting optical lattice clocks, like the thermal noise of the local oscillator. These are based on the principle of superradiant lasing of ultracold atoms inside a ``bad'' optical cavity. The operating principle has been demonstrated \cite{Bohnet2012SuperradiantLaserRb}, even on strontium's optical clock transition \cite{Norcia2016SuperradianceStrontiumClock}, but what is desirable is a high phase-space density (PSD) continuous atom source in order to run the clock steady-state \cite{Muniz2019ThompsonSuperradiantClockSPIE, iqClock}. Similarly, other cold atom sensors would benefit from continuous operation, like inertial sensors for navigation that could feature both absolute calibration and continuous measurement \cite{Jekeli2005NavigationErrorAnalysis}.

The development of atomic beams with high phase-space density has historically been closely tied with efforts to produce a steady-state atom laser \cite{Robins2013RevAtomLaser}, perhaps the ultimate source for many cold atom sensors. Previous work by \citet{Lahaye2004GuidedCollAtomBeam, Lahaye2005EvapBeam} demonstrated steady-state beams with a PSD of $10^{-7}$, by repeatedly outcoupling rubidium atoms from a MOT and evaporatively cooling them as they traversed a $\SI{4.5}{\meter}$ long magnetic waveguide. A chromium beam of $1 \times 10^{7} \, \mathrm{at}\,\si{\per\second}$ with a PSD of $3 \times 10^{-8}$ was produced in \cite{Aghajani-Talesh2010CrBeamPSD, Falkenau2011ContinuousLoadTrap}, and \citet{Knuffman2013IonBeamColdAtom} produced a caesium beam of $5 \times 10^{10} \, \mathrm{at} \, \si{\per\second}$ with a PSD of $4 \times 10^{-8}$ for a focused ion beam source.

Here we present a continuous $^{88}$Sr source delivering an atomic beam of $3 \times 10^{7} \, \mathrm{at}\, \si{\per\second}$ with a phase-space density of more than $10^{-4}$ in the moving frame. In order to enable an extremely low forward velocity of $\sim \SI{10}{\centi \meter \per \second}$, this beam is supported against gravity by an optical dipole guide. This source could feed an atom interferometer in continuous operation mode, in particular using the ${^1\mathrm{S}_0} \mathrm{-} {^3\mathrm{P}_0}$ clock transition in a magic wavelength guide \cite{Akatsuka2017AImagic, Hu2017AIwithSrClock}. Moreover, this high-PSD atomic beam could provide the gain medium for a steady-state superradiant laser \cite{Chen2005ActiveClock, Meiser2009MilliHzLaser, Norcia2016SuperradianceStrontiumClock}, and produce a clock laser with linewidth substantially narrower than the transition linewidth \cite{Debnath2018SuperradiantNarrowLinewidth}. Last, this beam could possibly be the source for a continuous atom laser \cite{Hagley1999QuasiCWatomLaser, Guerin2006GuidedAtomLaser, Mandonnet2000EvapCoolingBeam, Olson2014PressureDrivenEvap}.

This paper is structured as follows. Section~\ref{Sec:Method} describes the experimental setup and the various steps and methods necessary to produce the beam. In section~\ref{Sec:FiguresOfMerit}, we discuss figures of merit for characterizing cold atomic beams. We then present our measurement protocols and the results we obtain for two strontium isotopes in section~\ref{Sec:Results}. Lastly, in section~\ref{Sec:Discussion} we discuss possible applications for our continuous high phase-space density atomic beam and conclude.

\section{Experiment}
\label{Sec:Method}

\begin{figure}[tb]
\includegraphics[width=0.98\columnwidth]{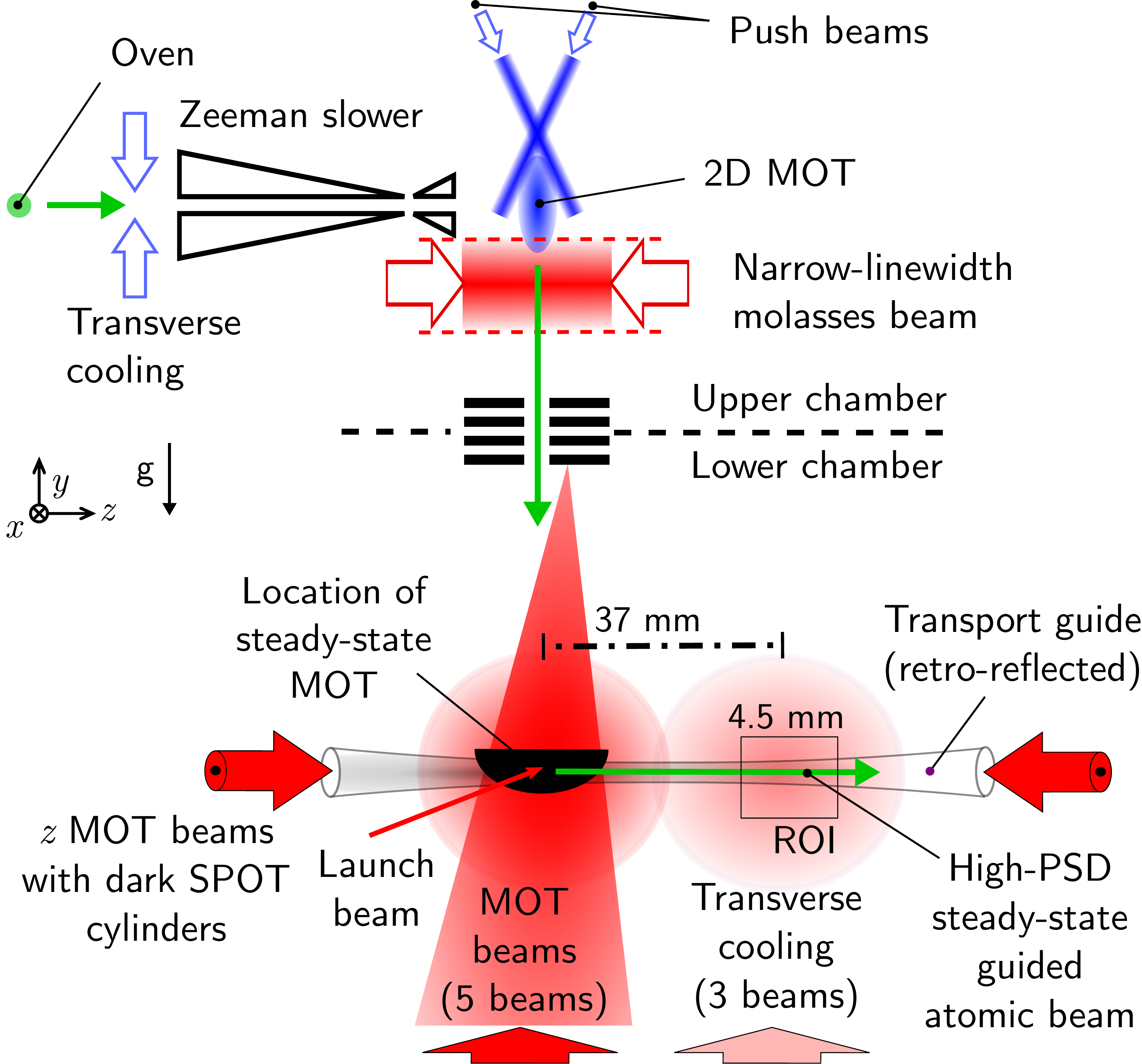}
\caption{\label{fig:Setup_schematic} Producing a high-PSD atomic beam. An oven emits strontium atoms that are transversely cooled and slowed by a Zeeman slower, whose output is caught by a 2D MOT, all operating on the ${^1\mathrm{S}_0} - {^1\mathrm{P}_1}$ transition (beams in blue). The ${^1\mathrm{S}_0} - {^3\mathrm{P}_1}$ transition (beams in red) is then used to radially cool the atoms by an optical molasses, before they fall under gravity to a baffled second chamber. Here atoms are recaptured into a steady-state 5-beam MOT, again operated on the ${^1\mathrm{S}_0} - {^3\mathrm{P}_1}$ transition. The MOT is overlapped with an optical guide into which atoms are launched with a $\sim \SI{10}{\centi\meter \per \second}$ mean velocity. This guided beam is radially cooled along the way by three low intensity molasses beams. The region of interest (ROI) where we characterize the atomic beam is located $\SI{37}{\milli\meter}$ from the MOT.}
\end{figure}

Our approach to produce a continuous, cold, bright atomic beam is based on flowing gaseous strontium through a series of spatially distributed laser cooling and guiding stages. Our scheme is illustrated in Fig.~\ref{fig:Setup_schematic}. The first stages are responsible for cooling atoms beginning with an $\sim \SI{800}{\kelvin}$ oven and finishing with a $\sim \SI{10}{\micro\kelvin}$ steady-state magneto-optical trap (MOT). We have previously reported on this MOT architecture in \cite{Bennetts2017HighPSDMOT}. We next outcouple atoms from the MOT into an optical dipole guide creating a bright, slow atomic beam. Transverse cooling of this atomic beam together with measures to detune and prevent MOT light from interacting with the beam are critical to achieve high performance. We characterize the resulting atomic beam at a location $\sim \SI{37}{\milli\meter}$ from the MOT. In the following section, we explain the details of this apparatus.

\textbf{Steady-state MOT atom source ---}
Beginning from an $\SI{800}{\kelvin}$ hot oven, atoms are transversely cooled, Zeeman slowed and captured by a 2D continuous magneto-optical trap (MOT) with ``push" beams, all operated on the $\SI{30}{\mega\hertz}$ linewidth ${^1\mathrm{S}_0} - {^1\mathrm{P}_1}$ transition. This first stage creates a beam with a flux of $\num{2.7e9} \, ^{88} \mathrm{Sr} \, \si{\per\second}$ at $\sim \SI{10}{\milli\kelvin}$ with a vertical downward velocity of a few meters per second. This beam is transversely cooled by a molasses operating on the $\SI{7.4}{\kilo\hertz}$ linewidth ${^1\mathrm{S}_0} - {^3\mathrm{P}_1}$ transition reducing the radial temperature to $\sim \SI{10}{\micro\kelvin}$ and allowing the beam to efficiently propagate to a second, lower chamber through a baffle. This baffle and dual chamber design was implemented to prevent ultracold atoms in the lower chamber from being heated by the continuously operating cooling light in the upper chamber. This can be critical depending on the time spent by the atoms in the lower chamber.

The falling atomic beam is captured by a steady-state 3D MOT operating on the $\SI{7.4}{\kilo\hertz}$ ${^1\mathrm{S}_0} - {^3\mathrm{P}_1}$ transition. The MOT geometry consists of five beams in orthogonal configuration. In the vertical axis, we shine a single MOT beam from below and rely on gravity to provide the downward restoring force. The MOT quadrupole magnetic field has gradients of $\{ 0.55, -0.32, -0.23 \} \, \si{\gauss \per \centi \meter}$ in the $\{ x, y, z \}$ axis, respectively. Atoms in the MOT typically have a temperature ranging from $5$ to $ \SI{30}{\micro \kelvin}$. The MOT laser detunings and intensities are adjusted to maximize the performance of the atomic beam at the region of interest (ROI, see Fig.~\ref{fig:Setup_schematic}) rather than the MOT itself.

In order to address atoms with Doppler shifts much larger than the $\SI{7.4}{\kilo \Hz}$ atomic linewidth, we use acousto-optic modulators to frequency broaden the MOT beams to a comb-like structure with a spacing $\delta \sim \SI{20}{\kilo\hertz}$ (corresponding to $2\mathrm{-}3 \, \times \Gamma_{{^1\mathrm{S}_0} - {^3\mathrm{P}_1}} / 2 \pi$). The detuning ranges $\left(\Delta_1; \delta; \Delta_2\right)$ are $ \left(-2.2; 0.015; -0.66\right) $, $ \left(-5.2; 0.02; -0.95\right) $ and $ \left(-2.2; 0.016; -0.82\right)\,\si{\mega \hertz}$ for the $x$, $y$ and $z$ axis, respectively. The power in each $\{ x, y, z \}$ axis MOT beam is $\{ 1.2, 10.8, 1.14\}\,\si{\milli \watt}$ and the $1/e^2$ beam diameter is ${\{ 47, 68, 48\}}\,\si{\milli \meter}$. The single beam in the $y$ axis is focused $\SI{22}{\centi \meter}$ above the MOT quadrupole magnetic field center and its $1/e^2$ diameter is $\sim \SI{35}{\milli \meter}$ at the MOT location.

The MOT beams along the $z$ axis provide confinement that can prevent the emission of an atomic beam along the $z$ direction. To mitigate this problem, we bore an $\SI{8}{\milli \meter}$ diameter hole in the center of the two mirrors directing the $z$ MOT beams from each side of the vacuum chamber, see Fig.~\ref{fig:Setup_schematic}. This allows the insertion of an extra pair of low intensity MOT beams down the $z$ axis of the MOT, strong enough to trap the cold MOT but weak compared to the other MOT beams that are optimized to capture hot incoming atoms. These additional beams fill up the holes entirely with a $1/e^2$ diameter of $\sim \SI{8}{\milli \meter}$, and they have a smaller detuning range of $\left(-1.25; 0.017; -0.85\right)\, \si{\mega \hertz}$ and a much lower power of $\SI{5}{\micro \watt}$. In addition to facilitating the outcoupling of a guided atomic beam the resulting MOT cloud is also elongated along the $z$ axis resulting in a better spatial overlap with the guide. Further details describing the steady-state 3D MOT can be found in \cite{Bennetts2017HighPSDMOT}. 

\textbf{Transport guide ---}
We continuously load atoms from the steady-state MOT into a ``transport'' guide, formed by an optical dipole beam overlapped with the MOT cloud and propagating along the $z$ axis. The guide is produced by focusing $\SI{12}{\watt}$ from a $\SI{1070}{\nano \metre}$ ytterbium fiber laser (IPG YLR-20-LP with $\SI{1.1}{\nano\metre}$ linewidth) to a $\SI{92}{\micro \meter}$ $1/e^2$ radius waist at the location of the MOT. In order to extend the guide length and improve the uniformity of the guide potential depth we retro-reflect the (incoherent) beam, focusing its second pass $\sim \SI{35}{\milli \metre}$ away from the MOT on the $z$ axis in the direction of the atomic beam propagation, with the same waist as in the first pass. The \SI{35}{\milli \meter} distance between focii is chosen to be on the same order as the Rayleigh length (\SI{25}{\milli \meter}) of both of these beams. By adapting the power of the retro-reflected beam with a polarizing beam splitter and a $\lambda/2$ waveplate, the potential landscape along the guide can be tuned and flattened. The effective trap depth at the MOT location is $\sim 35-\SI{40}{\micro\K}$, and $\sim \SI{25}{\micro\K}$ \SI{37}{\milli \metre} away, where the potential is flattened in the propagation direction, and where the radial trapping frequency is $\omega_r = 2 \pi \times \SI{185(10)}{\hertz}$. This deep, large volume trap at the MOT location improves the loading efficiency, and the flat potential for the final beam helps with further laser cooling stages by reducing light shift variations. The off-resonant scattering rate in the guide is negligible at $\sim \SI{0.1}{\Hz}$.

\textbf{Dark SPOT cylinder ---}
In our implementation, two MOT beams are overlapped and co-linear with the transport guide in the $z$ axis. It typically takes $\sim \SI{0.4}{\second}$ for atoms in the atomic beam to propagate the $\SI{37}{\milli\meter}$ from the MOT to the characterization location. Over such a long interaction time even a very small amount of resonant MOT light would be sufficient for the MOT restoring force to return atoms from the atomic beam to the MOT, thus devastating the beam's flux and temperature. Two factors mitigate this effect. Firstly the MOT light intensity is much lower in the \SI{8}{\milli\meter} core beam. Secondly the ${^3\mathrm{P}_1}$ $m_{J}=-1$ state experiences a $\SI{0.48}{\mega \Hz\per\cm}$ Zeeman shift due to the MOT's quadrupole magnetic field, quickly shifting the atoms propagating in the beam out of resonance with the MOT light. However, these measures alone are not sufficient.

\begin{figure}[tb]
\includegraphics[width=0.98\columnwidth]{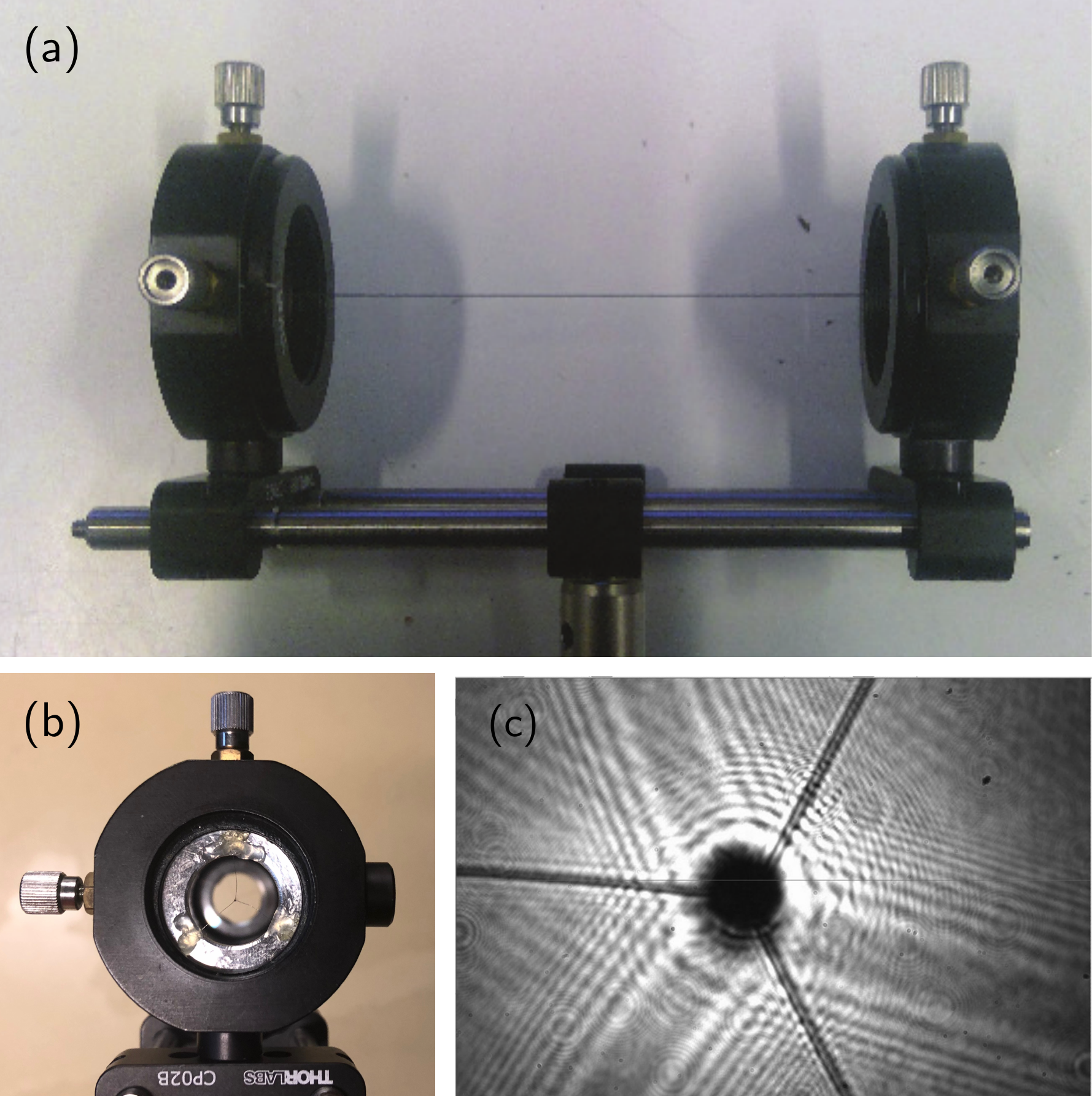}
\caption{\label{fig:Dark_SPOT_cylinder} MOT beam shadow casting structure. To protect the guided atomic beam from the co-propagating MOT light, two cylinders are imaged at the guide location, one for each MOT beam along the $z$ axis. (a) Side and (b) front view of an example of the shadow casting structure assembled. This structure consists of a capillary suspended using three threaded $\SI{50}{\micro\meter}$ diameter wires. (c) MOT light attenuation due to one cylinder, imaged at the plane corresponding to one of the cylinder's end. A dark cylinder achieves an attenuation of $\sim 30-40$ throughout its imaged length. The lines towards the edges are from the imaged suspension wires.}
\end{figure}

To further reduce the interaction between the MOT beams and the atomic beam we adapt the dark SPOT technique \cite{Ketterle1993FirstDarkSPOT}. Along both MOT beams in the $z$ axis, we image a $\SI{20}{\centi \meter}$-long, $\SI{600}{\micro \meter}$ outer diameter cylinder. By overlapping this shadow with the transport guide, we further darken the MOT region extending far beyond the characterization location \SI{37}{\milli\meter} from the MOT. The cylinder is assembled by suspending a stainless steel capillary within each inner $z$ axis MOT beam. The capillary has a $\SI{400}{\micro \meter}$ inner diameter through which three $\SI{50}{\micro \meter}$ diameter twisted wires pass. At each cylinder end the three thin wires are pulled taut triangularly sideways forming a tetrahedron shape and glued to an XY translator (Thorlabs, CXY1, 30 mm Cage XY Translator), see Fig.~\ref{fig:Dark_SPOT_cylinder}(a) and (b). The XY translators are used to precisely position each cylinder within a MOT beam. In front of each collimated MOT beam, we use a simple two-lens system in $f-2f-f$ configuration ($f = \SI{500}{\milli \meter}$) to image the cylinder's shadow onto the transport guide with a magnification of one.

We characterize the performance of these dark SPOT cylinders by imaging the shadow actually produced on the atoms, by collecting leakage light transmitted through MOT mirrors and imaging it onto a camera. By varying the image plane on the camera chip, we can select a specific object plane along the transport guide, as shown in the example of Fig.~\ref{fig:Dark_SPOT_cylinder}(c). Using this method, we checked the alignment of the dark volume across the atom's whole $\SI{37}{\milli \meter}$ travelling distance. We measure an attenuation of the MOT light by a factor $\sim 30-40$ along the center of the transport guide due to the dark SPOT cylinders. The darkness is ultimately limited by imperfections of the cylinder's surface and imaging, and by Poisson's spot from diffraction. Another MOT beam geometry, without a beam in the $z$ axis, could be envisioned, but some sort of dark slit would still be required to darken the region where the guide crosses out of the MOT beam.

\textbf{Launch beam ---}
With both the reduced MOT beam intensities on the $z$ axis and the effect of the dark SPOT cylinders, we observe outcoupling of atoms into the transport guide followed by propagation across $\SI{37}{\milli \meter}$. However, the atoms' speed is extremely low, dictated mainly by the MOT temperature, and outside the experimentalist's control. The resulting flux varies strongly, and any imperfection in the engineered darkness result in a beam that appears to stop at seemingly random places. Moreover, low propagation speeds render the beam more vulnerable to losses such as background gas collisions and off-resonant scattering from the transport guide. To remedy this situation, we add a ``launch'' laser beam resonant with the ${^1\mathrm{S}_0} - {^3\mathrm{P}_1}$ $\pi$ transition. This $\SI{250}{\micro \meter}$-waist beam shines $\SI{30}{\nano \watt}$ of light at the overlap between the MOT and transport guide, forming an angle with the guide of \SI{6}{\degree}, see Fig.~\ref{fig:Setup_schematic}. With the help of this launch beam, we can outcouple MOT atoms into the guide with a well-controlled mean velocity ranging from 8 to $\SI{25}{\centi\meter \per \second}$, see Section~\ref{Sec:Results}. During the remainder of this work, we typically operate with a launch beam intensity corresponding to a measured velocity of $8-\SI{9}{\centi\meter \per \second}$.

\textbf{Transverse cooling ---}
By applying transverse cooling with the ${^1\mathrm{S}_0} - {^3\mathrm{P}_1}$ $\pi$ transition to atoms propagating along the guide, we can both minimize the atomic beam's transverse temperature and optimize flux by removing evaporative losses. To this end, we place three single frequency laser beams with propagation axes perpendicular to the atomic beam, one \SI{36}{\milli\metre} $1/e^2$ diameter beam propagating upward along the $y$ axis, and a counter-propagating \SI{28.8}{\milli\metre} $1/e^2$ diameter beam pair along the $x$ axis, as shown in Fig.~\ref{fig:Setup_schematic}. These beams are centred around $\SI{35}{\milli\meter}$ from the MOT center and have powers of $\SI{1.4}{\micro \watt}$ and $\SI{6.75}{\micro \watt}$ for the horizontal and vertical axis, respectively. This gives a peak combined intensity of $\sim 0.5$ times the saturation parameter. These transverse cooling beams have a frequency $\SI{80}{\kilo\Hz}$ blue detuned from the ${^1\mathrm{S}_0} - {^3\mathrm{P}_1}$ $\pi$ transition for free atoms. Due to the differential light shift from the guide this corresponds to a $\sim \SI{200}{\kilo\Hz}$ red detuning for atoms passing along the center of the guide. With this transverse cooling, the radial temperature is ultimately lowered to about $\SI{1}{\micro \kelvin}$ and the flux is increased by a factor of $\sim 2$.

\section{Figures of merit - an overview}
\label{Sec:FiguresOfMerit}

Within the literature a variety of measures have been employed to characterize the performance of atomic beams, with each measure optimized for different applications. In this section we shall introduce and summarize these figures of merit as well as put them into context for applications such as interferometry, gain media for superradiant lasers and atom lasers.

Assuming a Gaussian density distribution of the atoms in the radial direction, the beam flux $\Phi$ can be represented by
\begin{equation}
\label{eq_BeamfluxGaussian}
\begin{split}
\Phi &=\int^{\infty}_{0} n_{0} \exp\left(-\frac{r^2}{2 \Delta r^2}\right)\,2\pi r dr \,{\bar{v}_z} \\
     &= 2 \pi\Delta r^2 \,n_{0}\, {\bar{v}_z} = \rho_L {\bar{v}_z}
\end{split}
,
\end{equation}

\noindent
with $\Delta r$ the root-mean-squared 1D spatial spread, $n_0$ the peak density, $\rho_L$ the linear density and ${\bar{v}_z}$ the mean longitudinal velocity. All these parameters can be directly measured on our experiment. The flux density is the flux per unit cross-section area, given by $\rho_{\Phi} = \Phi / \pi \Delta r^2$.

We also give the beam performance in terms of the gas phase-space density, usually employed for ultracold and quantum degenerate gases. The PSD is expressed in the moving frame as 
\begin{equation}
\label{eq_PSD}
\begin{split}
\rho_{\mathrm{PSD}}   &= n_0 \, \lambda_{\textrm{dB},r}^2 \, \lambda_{\textrm{dB},z}\\
                      &= n_0 \left( \frac{h}{\sqrt{2 \pi  \, m k_B T_{r}}}\right)^2 \, \frac{h}{\sqrt{2 \pi  \, m k_B T_{z}}}
\end{split}
,
\end{equation}

\noindent
where $h$ is the Planck constant, $k_B$ the Boltzmann constant, $m$ the atomic mass and $\lambda_{\textrm{dB},r/z}$ the thermal de Broglie wavelength associated with the 1D temperature $T_{r/z}$ in the radial/axial direction. Since we observe that the velocities follow Gaussian distributions, these effective 1D temperatures are directly related to the measured root-mean-squared 1D velocity spreads by the relations $\Delta v_{r,z} = \sqrt{k_B T_{r,z} / m}$. 

There are two ways in which we estimate the peak atom number density $n_0$. Firstly, we can use absorption imaging and fit a Gaussian profile to estimate the peak density. Alternatively, we may assume that the atom density inside the guide follows a Boltzmann distribution with radial temperature $T_r$. The density distribution then follows $n(r) = n_{0,\mathrm{therm}} \exp\left(-\frac{U(r)}{k_{B}T_{r}}\right)$, where $U(r)$ is the potential energy due to the transport guide. This is valid in the case of a gas in thermal equilibrium thanks to a high collision rate. For a guide that is deep compared to the radial temperature, its radial potential can be approximated by a harmonic oscillator potential with trapping frequency $\omega_r / 2 \pi$. Using eq.~(\ref{eq_BeamfluxGaussian}), the peak density in the thermalized case is then related to the linear density by $\rho_{L} =  n_{0,\mathrm{therm}} \, 2 \pi k_B T_ r / m \omega_r^2 $, and the expression of the phase-space density of eq.~(\ref{eq_PSD}) can be written as
\begin{equation}
\label{eq_PSDtherm}
\begin{split}
\rho_{\mathrm{PSD, therm}}   &= n_{0,\mathrm{therm}} \, \lambda_{\textrm{dB},r}^2 \, \lambda_{\textrm{dB},z} \\ 
                             &= \rho_{L} \left( \frac{h \omega_r}{ 2 \pi \, k_B T_r}\right)^2 \, \frac{h}{\sqrt{2 \pi  \, m k_B T_{z}}}
\end{split}
.
\end{equation}

\noindent
The temperatures $T_r$ and $T_z$, the trapping frequency $\omega_r / 2 \pi$ and the linear density $\rho_L$ are directly accessible from the experimental data.

Another quantity of interest for establishing the usefulness of a beam source is its brightness (or radiance). In the literature it is often expressed by the flux density divided by the solid angle $\Omega$ of the beam divergence, and it is of primary interest to characterize ion beams \cite{McClelland2016ReviewIonFromMOT}. Since our beam is strongly confined by the transport guide and since the axial speed is very low, the beam divergence can be negligible between, for example, two regions of interrogation in an interferometry scheme. The brightness is thus not the most suitable quantity to characterize our beam. We nonetheless provide it for completeness in the case of a cold atomic beam \cite{Lison1999HighBrillCsBeam, Glover2015CollimationBeamMolasse}
\begin{equation}
\label{eq_brightness}
\mathcal{R}  = \frac{\Phi}{\pi \Delta r^2 \, \Omega},
\end{equation}

\noindent
with $\Omega = \pi (\Delta v_{r} /  {\bar{v}_z})^2$. Similarly, the brilliance $\mathcal{B}$ of the beam is given by
\begin{equation}
\label{eq_brilliance}
\mathcal{B}  = \mathcal{R} \frac{{\bar{v}_z}}{\Delta v_{z}} = \frac{\Phi \, {\bar{v}_z}^3}{\pi^2 \Delta r^2 \, \Delta v_{z} \Delta v_{r}^2} ,
\end{equation}

Since our beam is guided, a better suited figure of merit is the velocity brightness $\mathcal{R}_v$, expressed as the flux per unit of beam area per three dimensional velocity spread,
\begin{equation}
\label{eq_velocitybrightness}
\mathcal{R}_v   = \frac{\Phi}{\pi \Delta r^2 \,  \Delta v_{z}   \Delta v_{r}^2} .
\end{equation}

\noindent
This expression is commonly used to characterize atomic beams for interferometry based precision measurement \cite{Treutlein2001HighBrightSourceFountain, Miossec2002MagnLens, Riis1990AtomFunnel, Chen2000RbFunnel} and to characterize atom lasers \cite{Robins2013RevAtomLaser, Bloch1999CWoutputAtomLaser, Robins2006PeakBrightAtomLaser}.

\section{Results}
\label{Sec:Results}

We now present the results from characterizing the atomic beam at a location $\SI{37}{\milli \meter}$ away from the MOT center, in a region out of the MOT laser beam and out of resonance with scattered light from these beams. We measure the axial mean velocity and velocity spread, and the radial velocity spread. We also measure the density, linear density and transverse spatial spread. From these we infer the beam flux, phase-space density and brightness. All these quantities are summarized in Table~\ref{tab:BeamResults}.

\begin{table}[tb]
\caption{Characteristics of the beam, for both $^{88}\mathrm{Sr}$ and $^{84}\mathrm{Sr}$, measured $\SI{37}{\milli \meter}$ away from the MOT. The symbols and their expressions are detailed in the main text. All uncertainties are taken as the standard deviation from the fitted data. The quantities ${\bar{v}_z}$ and $\Delta v_{z}$ could not be measured for $^{84}\mathrm{Sr}$ as the flux is not large enough. We assume the same values as for $^{88}\mathrm{Sr}$, scaled by the mass ratio.}
\label{tab:BeamResults}
\begin{center}
\begin{tabular*}{\columnwidth}{@{\extracolsep{\fill}}lcc} \hline
\noalign{\smallskip}
Parameter & $^{88}\mathrm{Sr}$ beam & $^{84}\mathrm{Sr}$ beam\\ \hline
\noalign{\smallskip} \hline \noalign{\smallskip}
Axial temperature $T_z$ [\si{\micro\kelvin}] & $29(2)$ & $29(2)$  \\ 
Rad. temperature $T_r$ [\si{\micro\kelvin}] & $0.89(4)$ & $2.0(1)$ \\ 
Axial velocity ${\bar{v}_z}$ [\si{\cm\per\second}] & $8.4(4)$ & $8.8(4)$ \\ 
Vel. spread $\Delta v_{z}$ [\si{\cm\per\second}] & $5.2(2)$ & $5.3(2)$ \\ 
Vel. spread $\Delta v_{r}$ [\si{\cm\per\second}] & $0.92(2)$ & $1.41(4)$ \\ 
Spatial spread $\Delta r$ [\si{\micro\meter}] & $23.3(4)$ & $19.7(1.0)$ \\ 
Lin. density $\rho_L$ [at~\si{\per\meter}] & $3.88(8) \times 10^{8}$ & $1.04(5) \times 10^{7}$  \\
Peak density $n_0$ [at~\si{\per\cubic\meter}] & $1.14(4) \times 10^{17}$ & $4.2(5) \times 10^{15}$  \\ 
Flux $\Phi$ [at~\si{\per\second}] & $3.25(14) \times 10^{7}$ & $9.1(6) \times 10^{5}$  \\ 
Flux dens. $\rho_{\Phi}$ [at~\si{\per\second \per \square \meter}] & $1.02(5) \times 10^{20}$ & $2.9(4) \times 10^{17}$  \\ 
Trap freq. $\omega_r / 2 \pi$ [\si{\hertz}] & $185(10)$ & $185(10)$ \\ 
Collision rate $\Gamma_{\mathrm{el}}$ [\si{\hertz}] & $3.5(2) \times 10^{-4}$ & $0.11(2)$  \\ 
PSD $\rho_{\mathrm{PSD}}$ & $1.5(2) \times 10^{-4}$ & $2.7(4) \times 10^{-6}$ \\ 
Alternate PSD $\rho_{\mathrm{PSD, therm}}$ & $1.3(2) \times 10^{-3}$ & $7.1(1.2) \times 10^{-6}$  \\
Bright. $\mathcal{R}$ [at~\si{\per \second \per \square\meter \per \steradian }] & $5.0(7)\times 10^{17}$ & $9(2) \times 10^{15}$ \\ 
Brill. $\mathcal{B}$ [at~\si{\per \second \per \square\meter \per \steradian }] & $2.5(5) \times 10^{18}$ & $5(1) \times 10^{16}$ \\ 
Vel. bright. $\mathcal{R}_v$ [at~\si{\square \second \per\meter\tothe{5}}] & $4.3(4) \times 10^{21}$ & $7(1) \times 10^{19}$ \\ 
\hline
\end{tabular*}
\end{center}
\end{table}

\textbf{Axial velocity and velocity spread ---}
Due to the extremely low temperatures reached in a strontium MOT operated on the narrow ${^1\mathrm{S}_0} - {^3\mathrm{P}_1}$ line, the resulting atomic beam can be extremely slow. This often welcome feature prevents us from characterizing the beam velocity and velocity spread by the conventional method of Doppler sensitive laser-induced fluorescence (LIF) \cite{Phillips1982ZeemanSlower}. In order to have a LIF signal with enough velocity resolution, the Doppler shift has to be large compared with the fluorescence light transition linewidth. The velocity of our atomic beam is on the order of $\SI{10}{\centi \meter \per \second}$, corresponding to a Doppler shift of only $\SI{217}{\kilo \Hz}$ for the $\SI{30}{\mega \Hz}$-wide ${^1\mathrm{S}_0} - {^1\mathrm{P}_1}$ transition, clearly an insufficient resolution. Alternatively, the weak $\SI{7.4}{\kilo \Hz}$-wide ${^1\mathrm{S}_0} - {^3\mathrm{P}_1}$ transition would give $\SI{145}{\kilo \Hz}$ of shift but insufficient fluorescence signal due to the low scattering rate.

\begin{figure}[tb]
\includegraphics[width=.98\columnwidth]{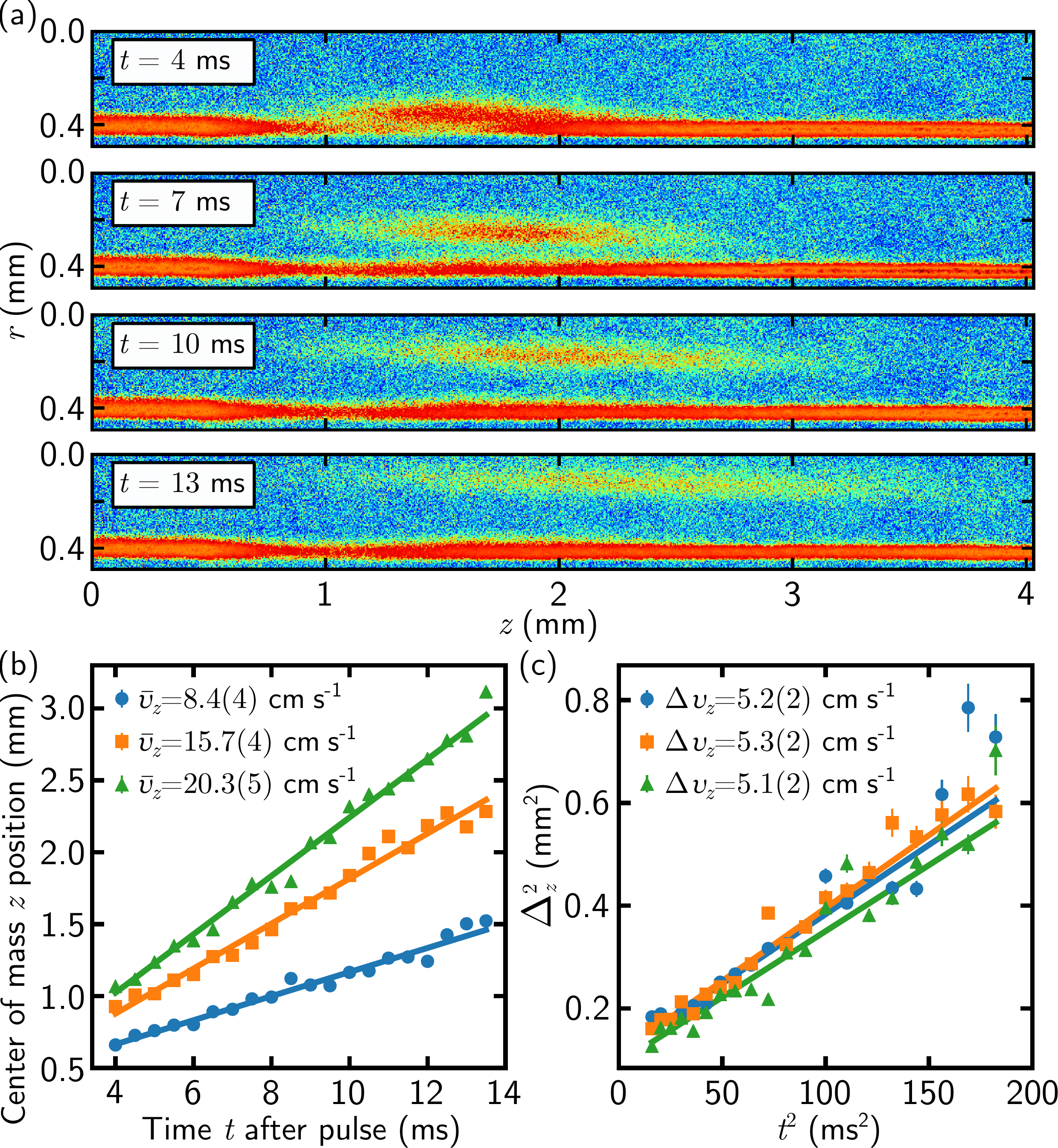}
\caption{\label{fig:Beam_axial_velocity_and_velocity_spread} Axial velocity and velocity spread of the atomic beam. (a) Steady-state atomic beam after having applied a $\SI{1}{\milli \second}$ pulse of light resonant with the ${^1\mathrm{S}_0} - {^1\mathrm{P}_1}$ transition. The ejected wavepacket is in free flight along $z$ for a varying time $t$, after which an absorption image is taken. (b) Center-of-mass position and (c) width $\Delta z$ of the ejected wavepacket, for three different intensities of the launch beam (see Section~\ref{Sec:Method}). Lines are linear fit from which we extract (b) the axial velocity ${\bar{v}_z}$ and (c) the velocity spread $\Delta v_{z}$. Error bars on data points in (b) and (c) show the standard deviation from Gaussian fits of the ejected wavepacket.}
\end{figure}

In order to measure the axial atomic beam velocity and velocity spread, we instead apply a $\SI{1}{\milli \second}$ pulse of light resonant with the ${^1\mathrm{S}_0} - {^1\mathrm{P}_1}$ transition, with a horizontal beam perpendicular to the guided atomic beam. This pulse ejects a packet of atoms out of the guide, see Fig.~\ref{fig:Beam_axial_velocity_and_velocity_spread}(a). We assume that spontaneously emitted photons are equally distributed in all directions during the ejection process, so that the $z$-axis mean velocity is not affected. We infer the mean velocity ${\bar{v}_z}$ and velocity spread $\Delta v_{z}$ by examining the evolution of the atom packet propagating alongside the transport guide. We perform this characterization for atoms in the transport guide located $\SI{37}{\milli \meter}$ away from the MOT, within the $\sim \SI{4}{\milli \meter}$ long region of interest (ROI) of our imaging setup. Thanks to our tunable transport guide architecture (see Sec.~\ref{Sec:Method}), the potential landscape along the guide axis is essentially flat across the ROI. We therefore assume the beam velocity to be constant throughout the ROI. We measure the velocity for several launch beam intensities, and confirm that we can adjust the mean velocity within the range of $8$ to $\SI{25}{\centi\meter \per \second}$, see Fig.~\ref{fig:Beam_axial_velocity_and_velocity_spread}{(b)}. We observe that for all launch beam intensities the atomic beam has a similar velocity spread of $\Delta v_{z} = \SI{5.2(2)}{\centi\meter \per \second}$ (see Fig.~\ref{fig:Beam_axial_velocity_and_velocity_spread}(c)). This velocity spread corresponds to a 1D temperature of $T_z = \SI{29(2)}{\micro \kelvin}$. By looking at the momentum imparted to the atomic packet by the ejection pulse in the radial direction, we check that the heating due to this pulse is negligible compared to the measured axial temperature.

\begin{figure}[tb]
\centering
\includegraphics[width=.98\columnwidth]{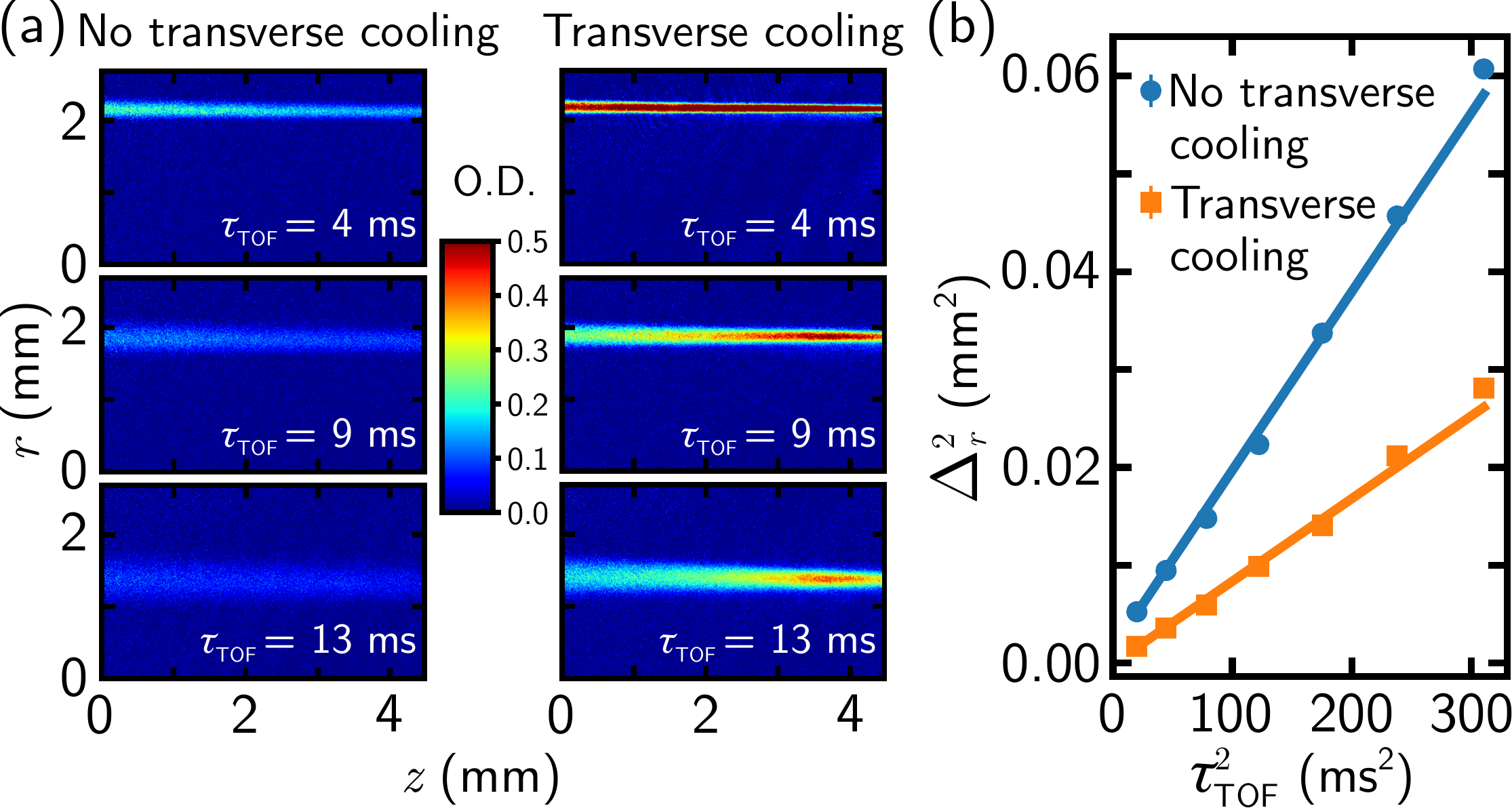}
\caption{\label{fig:Beam_radial_velocity_spread} Effect of transverse cooling on the beam radial velocity spread. (a) Absorption images of the steady-state atomic beam, taken after switching the transport guide and laser cooling beams off, and letting the cloud expand for various time of flights $\uptau_{\mathrm{TOF}}$. The beam is shown both without (left) and with (right) transverse cooling applied (see Section~\ref{Sec:Method}). For the sake of clarity, we set the maximum optical density (O.D.) to 0.5 on the color scale, which means that some pictures are saturated. (b) Expansion of the beam size $\Delta_r$ in free flight, from which we extract the radial velocity spread. Lines are linear fits, giving $\Delta v_r = \SI{1.36(2)}{\centi \meter \per \second}$ without and $\Delta v_r = \SI{0.92(2)}{\centi \meter \per \second}$ with transverse cooling. This corresponds to a radial temperature $T_r = \SI{1.93(5)}{\micro \kelvin}$ and $T_r = \SI{0.89(4)}{\micro \kelvin}$, respectively. The small error bars in (b) give the standard deviation from Gaussian fits of density profiles integrated along the whole ROI.}
\end{figure}

\textbf{Radial velocity spread ---}
We measure the radial velocity spread in a more conventional way, by switching off the transport guide and measuring the atomic beam size after ballistic expansion, see Fig.~\ref{fig:Beam_radial_velocity_spread}(a). We obtain a radial temperature of $\SI{1.93(5)}{\micro \kelvin}$ averaged over the entire ROI. With the addition of transverse cooling light (see Section~\ref{Sec:Method}), the flux increases by a factor of $\sim 2$ and the radial temperature reduces to $T_r = \SI{0.89(4)}{\micro \kelvin}$, corresponding to a velocity spread of $\Delta v_{r} = \SI{0.92(2)}{\centi\meter \per \second}$, see Fig.~\ref{fig:Beam_radial_velocity_spread}(b).

Within the ROI, we observe a mild variation of both the density and the radial velocity spread along the transport guide axis. More precisely, as the atoms propagate, both the flux and the radial temperature reduce. Without transverse cooling light, this can be explained by radial losses, which could be enhanced by eventual corrugations of the guide potential that transfer momentum from the axial to the radial direction. When the transverse cooling light is applied, the radial temperature slowly decreases, as expected, with the atoms' travelling time. The losses are thus strongly reduced, leading to a slower decrease of the flux.

\textbf{Density, flux, brightness ---}
Our imaging system resolution of $\SI{4.5}{\micro \meter}$ is sufficient to image the beam density profile by absorption imaging (with a negligible $\SI{0.1}{\milli \second}$ time of flight) while staying within the dynamic range of the camera. The linear density $\rho_L$ can be estimated by integrating the atom number over the radial direction and along the full length of the ROI. We fit the beam averaged along the propagation axis to the Gaussian profile of eq.~(\ref{eq_BeamfluxGaussian}), in order to estimate the peak density $n_0$ and the spatial spread $\Delta r$. From all the measured quantities and the expressions given in Section~\ref{Sec:FiguresOfMerit}, we extract the results summarized in Table~\ref{tab:BeamResults} for both $^{88}\mathrm{Sr}$ and $^{84}\mathrm{Sr}$ isotopes. From the measured flux captured by the MOT and the flux observed at the end of the guide, we estimate the transfer efficiency from the MOT to the beam to be $\sim \SI{30}{\percent}$.

\textbf{Thermalization ---}
As shown in Table~\ref{tab:BeamResults}, there is a clear discrepancy between the PSD values of $\rho_{\mathrm{PSD}}$ and the alternate $\rho_{\mathrm{PSD, therm}}$, with the latter being higher by a factor of $\sim 10$ for $^{88}\mathrm{Sr}$ and $\sim 3$ for $^{84}\mathrm{Sr}$. The expression for $\rho_{\mathrm{PSD}}$ estimates the density based on the measured atomic beam radius rather than assuming a Boltzmann distribution in a harmonic trap. We can therefore understand this discrepancy from the absence of thermalization, which can be seen from the differences between axial and radial temperatures. Without a thermalized sample, $\rho_{\mathrm{PSD, therm}}$ is not reliable so we shall keep the smaller value for PSD given by $\rho_{\mathrm{PSD}}$. 

Following the work of \cite{Anderlini2005CollAtomMixt}, we can estimate the elastic collision rate within the beam. From our measurements of the anisotropic density and velocity distributions, we can estimate the elastic collision rates to be $\Gamma_{\mathrm{el,88}} = \SI{3.5(2)e-4}{\hertz}$ and $\Gamma_{\mathrm{el,84}} = \SI{0.11(2)}{\hertz}$. Given the propagation time from the MOT of less than $\SI{0.5}{\second}$, the scattering rates for both isotopes are insufficient to thermalize the atoms.

Let us note that, despite the much higher density, the collision rate for $^{88}\mathrm{Sr}$ is $\sim 300$ times smaller than for $^{84}\mathrm{Sr}$, due to the scattering length $a_{\mathrm{scat},88} = -1.4 \, a_0$ being much smaller than $a_{\mathrm{scat},84} = 122.7 \, a_0$, with $a_0$ the Bohr radius. These rates can be expressed as $\Gamma_{\mathrm{el}} = n_0 \sigma_{\mathrm{cross}} v_\mathrm{coll}$, where $\sigma_{\mathrm{cross}} = 8 \pi a_{\mathrm{scat}}^2$ is the elastic cross-section and $v_\mathrm{coll} = \sqrt{2 k_B T_{\mathrm{eff}} / \pi m}$ is the mean relative velocity, for an effective isotropic temperature $T_{\mathrm{eff,88}} \sim \SI{8}{\micro \kelvin}$ and $T_{\mathrm{eff,84}} \sim \SI{10}{\micro \kelvin}$.

\section{Discussion}
\label{Sec:Discussion}

The steady-state beam we demonstrate here has a phase-space density around three orders of magnitude higher than any previous steady-state atomic beam. In fact, the velocity brightness of $\mathcal{R}_v=\num{4.3(4)e21}$ for $^{88}\mathrm{Sr}$ is approaching what has been reached by pulsed quasi-CW atom lasers \cite{Robins2013RevAtomLaser, Robins2006PeakBrightAtomLaser}, for example $\mathcal{R}_v=\num{2e24}$ in Ref.~\cite{Bloch1999CWoutputAtomLaser}.

An immediate application for such a system might be continuously replenishing the gain medium of a steady-state superradiant active clock. There has been significant interest in the development of active optical clocks in recent years \cite{Chen2005ActiveClock, Meiser2009MilliHzLaser}. There have been pulsed demonstrations on the strontium clock transition  \cite{Norcia2016SuperradianceStrontiumClock} and a great deal of theoretical work \cite{Meiser2010PRAHolland, Kazakov2017Schumm, Debnath2018SuperradiantNarrowLinewidth, Zhang2019Molmer, Schaffer2019Thompsen, Hotter2019Ritsch}, but any active clock requires replenishment of the atoms used for the gain medium. Modelling suggests for the ${^1\mathrm{S}_0} \mathrm{-} {^3\mathrm{P}_0}$ clock transition that an ideal atomic source would consist of a guided continuous beam with a flux of $> \num{1e7} \, ^{88}\mathrm{Sr} \, \si{\per\second}$ and a velocity of $\sim \SI{10}{\centi \meter\per\second}$, criteria fulfilled by the beam we have demonstrated.

Another interesting application is steady-state interferometry schemes that aim to operate in a continuous mode, eliminating the Dick effect. This might be particularly important for long interrogation times, for example making use of the ${^1\mathrm{S}_0} \mathrm{-} {^3\mathrm{P}_0}$ transition \cite{Akatsuka2017AImagic, Hu2017AIwithSrClock} that has been proposed for gravitational wave detectors \cite{Hogan2011GWsensorAGISLEO}, although most likely an additional cooling stage is necessary for interferometry applications.

While the low axial velocity is ideal for some applications, it reduces both the brightness and brilliance figures of merit typically used for applications such as milling with ion beams. This, along with the reduced flux compared to other systems \cite{McClelland2016ReviewIonFromMOT}, would likely limit direct use in applications such as high current ion beam sources or high rate doping. However, it can be a good starting point for ion beam sources where the goal is to achieve the highest resolution. Similarly, this continuous, low spread beam is a good source for high repetition rate deterministic single ion sources \cite{Ates2013DeterminIonBlockade, Sahin2017DeterminIonHighRate}.

For many applications the ultimate atomic beam source would be a steady-state continuous atom laser and, in the past, efforts towards this goal have produced the highest phase-space density beams reported \cite{Lahaye2004GuidedCollAtomBeam, Lahaye2005EvapBeam}. The work here represents a major step towards this goal even though it is preliminary in many aspects. A better control of the axial beam velocity is possible in several ways, one of which has been recently demonstrated in our group \cite{Chen2019SOLD}. By reducing the velocity and increasing the density, it might be possible with the $^{84}\mathrm{Sr}$ isotope to increase the elastic collision rate $\Gamma_{\mathrm{el}}$ and reach a regime where collisions dominate \cite{Lahaye2004GuidedCollAtomBeam, Olson2014PressureDrivenEvap}. This would enable evaporative cooling along the dipole guide \cite{Lahaye2005EvapBeam, Mandonnet2000EvapCoolingBeam, Olson2006WaveguideEvap}, which would increase the beam PSD and hopefully produce a continuous atom laser \cite{Robins2013RevAtomLaser, Bloch1999CWoutputAtomLaser}.

To summarize, we have demonstrated a continuous guided atomic beam of $^{88}$Sr with a phase-space density more than three orders of magnitude higher than in previously reported systems. Our beam has an extremely low mean velocity ${\bar{v}_z}= \SI{8.4(4)}{\cm\per\second}$, radial spatial spread $\Delta r = \SI{23.3(4)}{\micro\meter}$ and radial velocity spread $\Delta v_{r} = \SI{0.92(2)}{\cm\per\second}$. This corresponds to a radial temperature of just \SI{0.89(4)}{\micro\kelvin}. The beam flux is $\Phi = \num{3.25(14)e7} \, \mathrm{at} \, \si{\per\second}$ and it reaches a PSD $\rho_{\mathrm{PSD}} = \num{1.5(2)e-4}$. Using the \SI{0.56}{\percent} abundant $^{84}$Sr isotope, we obtain a reduced phase-space density, but the higher scattering length means that our beam is approaching the collisionally dense regime, where evaporative cooling can be used to rapidly improve phase-space density. This represents a significant step towards the demonstration of a steady-state atom laser. Moreover, this beam is likely to find immediate application in efforts to demonstrate a steady-state superradiant active optical clock. A beam with such output performance could fulfill the demands of other applications requiring both ultracold atoms and uninterrupted operation, such as continuous atom interferometers, clocks, ion sources and steady-state atom lasers.

\begin{acknowledgments}
We thank Andrea Bertoldi for careful reading of the manuscript and providing insightful comments. We thank the Netherlands Organisation for Scientific Research (NWO) for funding through Vici grant No. 680-47-619 and the European Research Council (ERC) for funding under Project No. 615117 QuantStro. This project has received funding from the European Union’s Horizon 2020 research and innovation program under grant agreement No 820404 (iqClock project). B.P. thanks the NWO for funding through Veni grant No. 680-47-438. C.-C. C. thanks the Ministry of Education of the Republic of China (Taiwan) for a MOE Technologies Incubation Scholarship.
\end{acknowledgments}


%

\onecolumngrid

\end{document}